\documentclass{aa}  

\usepackage{graphicx}
\usepackage{txfonts}
\usepackage{multirow}
\usepackage{caption}
\usepackage{amsmath}
\usepackage{chemformula}
\usepackage[version=4]{mhchem}
\usepackage{natbib}
\usepackage{tabularx}

\begin{document}

   \title{Neon is an inhibitor of CO hydrogenation in pre-stellar core conditions}

   \author{B. Husquinet\inst{1,2}, J. Vitorino\inst{1}, O. Sipil{\"a}\inst{2}, P. Caselli\inst{2}
          \and
          F. Dulieu\inst{1}
          }

   \institute{CY Cergy Paris Université, Observatoire de Paris, Université PSL, Sorbonne Université, Université Paris Cité, CNRS, LIRA, F-95000 Cergy, France\\
   \email{francois.dulieu@cyu.fr}
         \and
             Center for Astrochemical Studies, Max-Planck-Institut für Extraterrestrische Physik (MPE), Gie\ss enbachstr. 1, D-85741 Garching, Germany\\
             \email{bhusquin@mpe.mpg.de}
             }

   \date{Received 07 March 2025 / Accepted 09 September 2025}

  \abstract
   {Neon (Ne) is the fifth most abundant element in the Universe. Because it is chemically inert, it has never been considered in astrochemical models that studied molecular evolution. In the cold dark environments of pre-stellar cores, where the temperatures are below 10\,K, Ne can condense onto the surface of interstellar grains. This might affect the formation of molecules.}
   {We investigated the effect of Ne on the production of formaldehyde (\ce{H2CO}) and methanol (\ce{CH3OH}) through carbon monoxide (CO) hydrogenation on different cold surfaces. We highlight its role in conditions corresponding to pre-stellar cores.}
   {In an ultra-high vacuum system, we conducted two types of experiments. The first experiment involved the co-deposition of CO and H atoms with or without Ne. The products were analysed using a quadrupole mass spectrometer. The second experiment involved depositing a monolayer of CO and separately a monolayer of Ne (or vice versa), followed by bombarding the layers with hydrogen atoms. We measured the evolution of the CO layer through Fourier-transform infrared absorption spectroscopy. Additionally, we used a gas-grain chemical code to simulate a pre-stellar core and determine where Ne can affect the chemistry.}
   {The presence of Ne on the surface significantly inhibits CO hydrogenation at temperatures below 12\,K. In the co-deposition experiments, we observed a 38\,\% decrease in the \ce{H2CO} production at 11\,K when the quantity of Ne in the mixture was lower than a monolayer. At 10\,K and with one monolayer in the mixture, the production decreased to 77\,\%, and it reached 91\,\% for a few monolayers of Ne in the mixture at 9\,K. While the decrease in \ce{CH3OH} formation is still notable, it is less pronounced: 43\,\% at 11\,K, 61\,\% at 10\,K, and 77\,\% at 9\,K. Experiments with stacked layers revealed that the CO layer decay varies slightly when the Ne layer is positioned above or below it. This observation indicates that Ne and CO create a mixture in which Ne can diffuse and stabilize at the surface, which isolates CO molecules from the accreting H atoms. Gas-grain chemical modelling showed that the first layer of Ne condenses in the central area of a pre-stellar core, typically within 5000\,AU, where CO molecules completely freeze out onto grains.}
   {Ne inhibits the hydrogenation process in the very central part of pre-stellar cores, and in general, where temperatures drop below 9\,K and the density increases above 10$^4$\,cm$^{-3}$.}


    \authorrunning{B. Husquinet et al.}

   \maketitle

\section{Introduction}

The first phases of star formation are observed in dense cores, which are extremely cold regions of the interstellar medium.
 In particular, in pre-stellar cores with central temperatures of about 7\,K \citep{crapsi2007observing,pagani2007depletion,launhardt2013earliest} and central densities higher than 10$^5$\,cm$^{-3}$\citep{keto2008different}, most of the chemical species in the gas phase stick to the surface of the dust grains, where they form icy mantles \citep{bergin2007cold}. In these environments, some molecules, such as CO, have high depletion factors that suggest significant freeze-out onto dust grains \citep{caselli1999co,bacmann2002degree,pagani2012method}. Within the central 2000\,AU of a pre-stellar core, it has been observed that 99.9\,\% of all chemical species heavier than He undergo freeze-out \citep{caselli2022central}. 

The formation of molecules on ice is a complex process that involves gaseous species that accrete onto interstellar dust grains. These grains act as catalysts and concentrators, which enable reactions between different species that migrate on their surfaces to form molecules of increasing complexity \citep{tielens1982model,tsuge2023radical}. The most abundant molecule in the Universe, \ce{H2}, can only be formed efficiently on the surface of grains \citep{gould1963interstellar,hollenbach1971surface,cazaux2002molecular,wakelam2017h2,grieco2023enhanced}. Simple molecules such as \ce{CO2}, \ce{H2O}, \ce{CH4}, \ce{CH3OH}, and \ce{NH3}, which have been observed in various environments in solid state, are formed at low temperatures on grain surfaces \citep{boogert2015observations}. Similarly, complex organic molecules (COMs), whose presence can be explained by solid-phase formation, are also formed at low temperatures \citep{herbst2009complex,butscher2015formation,fedoseev2015experimental,Dulieu2019formamide,ioppolo2021non}. Some species on the surface can prevent other species from reacting as well, however. One of these, neon, can play an important role.

Neon (Ne) is the fifth most abundant element in the Universe, following hydrogen, helium, oxygen, and carbon \citep{cameron1973abundances,lodders2010solar,nicholls2017abundance}. Even though often neglected in astrochemistry, possibly because it is hardly observable, it is more abundant than nitrogen, for example.  Despite its high abundance, it is not included in molecular evolution models because it is a stable element that belongs to the noble gases, and therefore, it does not recombine with itself or other species. Furthermore, the depletion of Ne in the gas phase towards the solid phase is not expected \citep{sofia1994abundant,brinchmann2013estimating}, in contrast with oxygen. This makes the O/Ne ratio a highly effective indicator of the chemical evolution of galaxies \citep{dors2013optical,armah2021chemical}.
Laboratory studies have shown, however, that Ne sticks in small quantities to the surface of grains below 10\,K \citep{schlichting1993techniques}. This indicates a need for an evaluation and consideration of the presence of Ne on the ice in pre-stellar cores and for studying its impact on the formation of molecules on ice. 

We present the first results of our experimental study of the impact of Ne on the formation of formaldehyde (\ce{H2CO}) and methanol (\ce{CH3OH}) induced by the hydrogenation of carbon monoxide (CO). The paper is structured as follows: We briefly describe the experimental methods in Section\,\ref{sec: Experimental method}, the results and discussion in Section\,\ref{sec: Results and Discussion}, and we discuss the astronomical implications in Section\,\ref{sec: Astrophysical implications}. Finally, a summary is provided in Section\,\ref{sec: Conclusion}.

\section{Experimental method} \label{sec: Experimental method}

The experiments were carried out using VENUS (VErs de NoUvelles Synthèse) setup, an ultra-high vacuum (UHV) multi-beam apparatus at the LIRA-CY laboratory. Details of the experimental setup are given in Appendix\,A, notably a sketch of the apparatus, and in \cite{congiu2020new}. We only discuss the experimental method we used for the various experiments we carried out for this study. Briefly, frozen molecular films, hereafter referred to as ices, were grown in an UHV chamber with a base pressure of $2\times10^{-10}$\,hPa on a gold-coated copper substrate at temperatures between 9 and 12\,K. Compact amorphous solid water (cASW) can also be grown on gold before the deposition of other gases to create a water-ice substrate of at least 10\,ML (ML is a monolayer with a surface density unit of $1\times10^{15}$ particles per cm$^2$. It corresponds to approximately one layer of adsorbate). In this case, water molecules were deposited via the background vapour inlet while the substrate was held at 110\,K \citep{congiu2020new}. Ices were grown by gas deposition using collimated effusive beams aimed at the substrate with an angle close to the normal incidence. The ice evolution was monitored by Fourier-transform infrared reflection and absorption spectroscopy (FT-RAIRS) during the deposition with a spectral resolution of $R=2$\,cm$^{-1}$ and a rate of one spectrum every 128\,s. After the deposition, a temperature-programmed desorption (TPD) with a thermal ramp of $\beta=0.2$\,K/s was performed using the quadrupole mass spectrometer (QMS) placed in front of the sample.

We deposited three different species using separated beam lines: CO, Ne, and atomic hydrogen (H). Hydrogen atoms were generated by dissociating \ce{H2} molecules in a microwave plasma source, achieving a dissociation efficiency of over $50\,\%$ at the sample. The CO molecular beam was calibrated as described by \cite{kruczkiewicz2024comprehensive}, with the first monolayer reached in $9\pm1$\,minutes, resulting in a flux of $\Phi_{\ce{CO}}=(1.85\pm0.19)\times10^{12}$\,cm$^{-2}$s$^{-1}$. The Ne beam was calibrated based on the CO reference, with a flux of $\Phi_{\ce{Ne}}=(2.38\pm0.3)\times10^{12}$\,cm$^{-2}$s$^{-1}$, allowing the first ML to be reached in $7\pm1$\,minutes.

Two types of experiments were performed. The first type, called co-deposition, involved depositing CO and H or CO, Ne, and H simultaneously for 60\,minutes on a substrate held at a constant temperature set between 9 and 12\,K. We refer to the co-deposition of CO and H as \{CO+H\} and to the co-deposition of CO, Ne, and H as \{CO+H+Ne\}. The second type of experiments involved the H atom exposure of 1\,ML thick CO or Ne ice for 75\,minutes with a substrate temperature of 9\,K.

The surface coverage $N_X$ of molecules remaining on the surface, obtained via the TPD data, is calculated using the following equation:
\begin{equation}
    N_X = \frac{A_X}{A_{CO}^{ML}}\frac{\sigma^{Tot}_{CO}}{\sigma^{Tot}_X} \;,
    \label{eq: Nx}
\end{equation}
where $A_X$ corresponds to the total area of molecule $X$ in the TPD profile (the cracking pattern is taken into account; \citealt{vitorino2024sulphur}), while $A_{CO}^{ML}$ is the reference area for one CO monolayer. To correct for ionization effects, we considered for each molecule the total ionization cross section per electron impact for an electron energy of 30\,eV, represented as $\sigma^{Tot}_X$. For reference, the total cross section of CO is $\sigma^{Tot}_{\ce{CO}}=1.296\,\AA^2$ \citep{hwang1996new}. For \ce{H2CO}, the total cross section is $\sigma^{Tot}_{\ce{H2CO}}=2.74\,\AA^2$ \citep{vinodkumar2011electron}, while $\sigma^{Tot}_{\ce{CH3OH}}=3.263\,\AA^2$ for \ce{CH3OH} \citep{kumar2019electron}. For Ne, $\sigma^{Tot}_{\ce{Ne}}=0.0805\,\AA^2$ \citep{rejoub2002determination}.
We neglected the filtering effect of the mass spectrometer, which is almost constant for masses between 20\,a.m.u (Ne) and 32\,a.m.u (\ce{CH3OH}, a product of CO hydrogenation). 

\section{Results and discussion} \label{sec: Results and Discussion}

\begin{figure*}
    \centering
    \includegraphics[width=\textwidth]{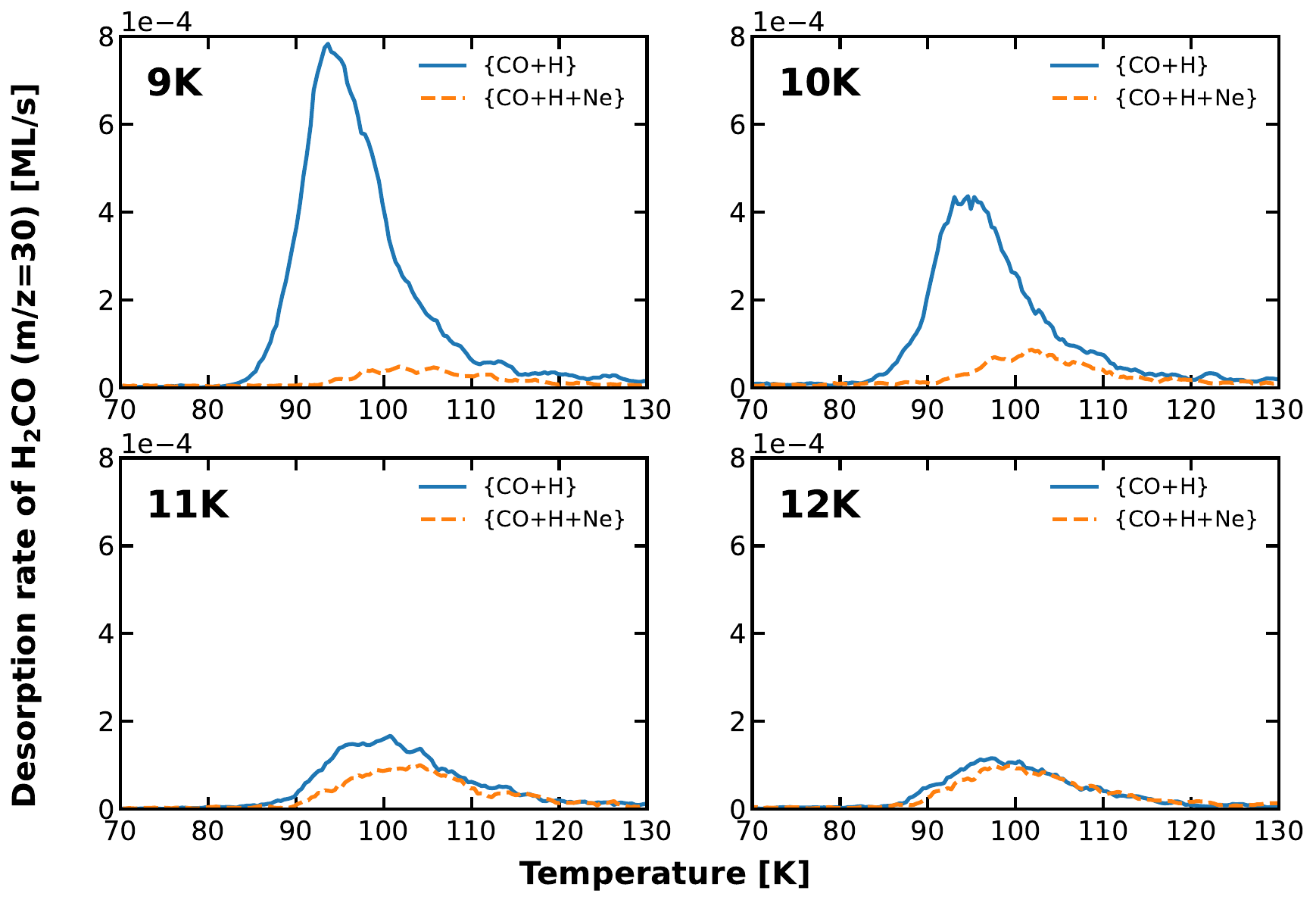}
    \caption{TPD curves for \ce{H2CO} ($m/z$=30) on a gold substrate at four different temperatures: 9, 10, 11 and 12\,K. The blue line represents the co-deposition of CO and H, and the dashed orange line represents the co-deposition of CO, H, and Ne.}
    \label{figNe: TPD H2CO}
\end{figure*}

\begin{figure*}
    \centering
    \includegraphics[width=\textwidth]{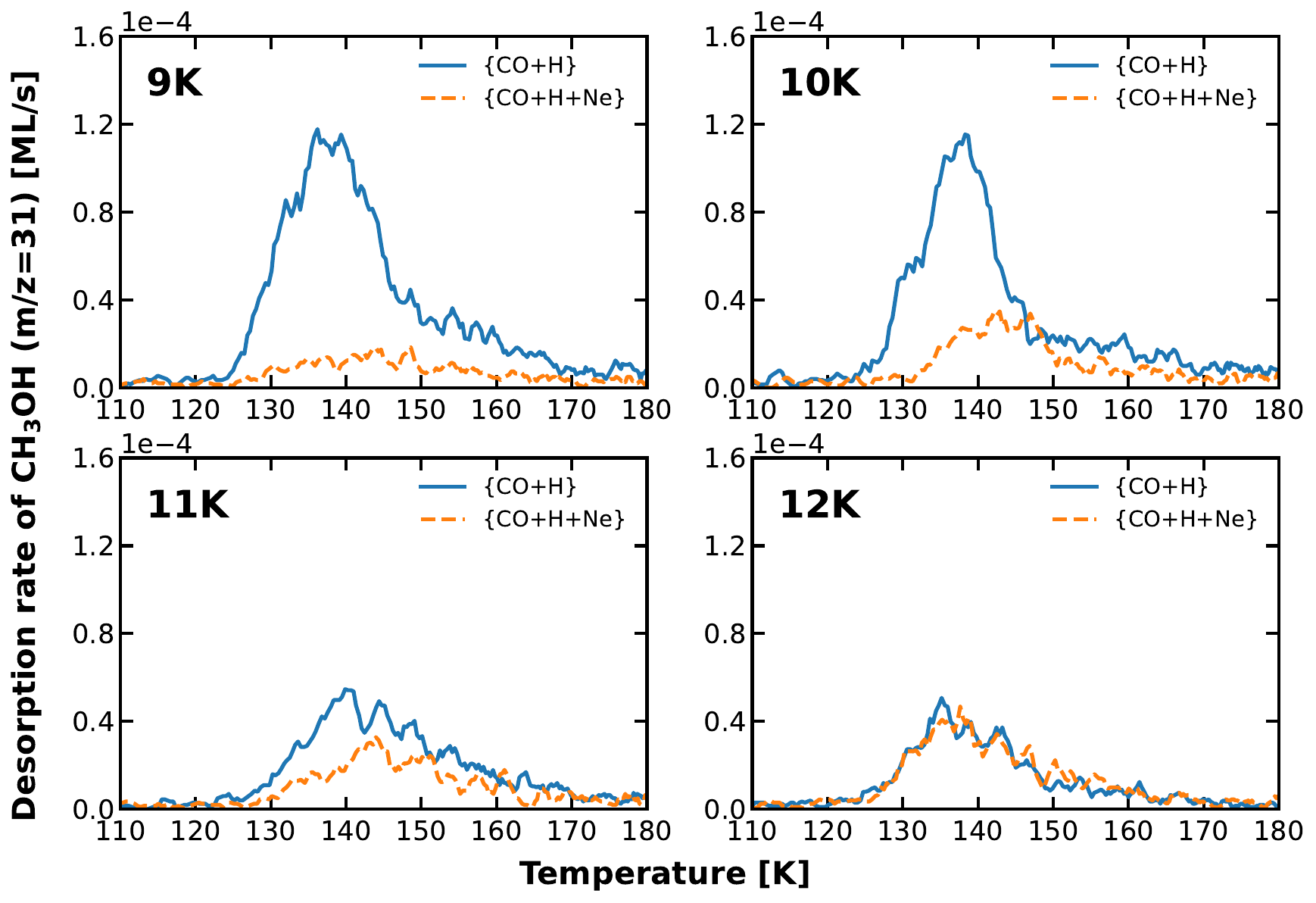}
    \caption{As in Fig.\,\ref{figNe: TPD H2CO}, but for \ce{CH3OH} ($m/z$=31).}
    \label{figNe: TPD CH3OH}
\end{figure*}

Figure\,\ref{figNe: TPD H2CO} shows TPD curves representing \ce{H2CO} ($m/z$=30) when formed via the \{CO+H\} co-deposition (blue lines) or via the \{CO+Ne+H\} co-deposition (dashed orange lines). In all cases, the species were deposited during 60\,minutes, but at different surface temperatures. At 12\,K (bottom right panel of Figure \ref{figNe: TPD H2CO}), the two TPD curves are almost indistinguishable and the surface coverage of \ce{H2CO} measured are the same within the precision limit of our experiment. As the surface temperature decreased, the difference between the two TPD curves increased. Less \ce{H2CO} was detected when Ne was present on the surface. The greatest difference was seen at 9\,K (top left panel of Figure \ref{figNe: TPD H2CO}), where we measured 11\,times more \ce{H2CO} for \{CO+H\} than for \{CO+H+Ne\}. A similar phenomenon was observed for \ce{CH3OH}: As reported in Table\,\ref{tab: Rapport mesure} and represented in Figure\,\ref{figNe: TPD CH3OH}, less \ce{CH3OH} was formed when neon was present on the surface. The fraction of \ce{CH3OH} measured in the presence of Ne also decreased the lower the temperature. We conclude that Ne significantly affects the hydrogenation process when the surface temperature is below 11\,K. 

The binding energy of a pure neon matrix is $E_b$ = 267\,K \citep{schlichting1993techniques}, which would correspond in our case to multi-layers of Ne. The binding energy of the first layers is slightly higher and dependent on the coverage (see below). The binding energy is not high enough, however, and the thermal desorption can therefore not be neglected. The desorption rate increases with the coverage, so that it becomes greater than the accretion rate. This implies a low effective accretion of Ne for temperatures close to or above 12\,K. Ne might accrete on the surface, but the desorption rate is faster than the accretion rate. When Ne is exposed to the surface, only a very low coverage is therefore achieved at 12\,K. In contrast, the effective coverage is larger when the surface temperature is low. 

\begin{figure}
    \centering
    \includegraphics[width=\columnwidth]{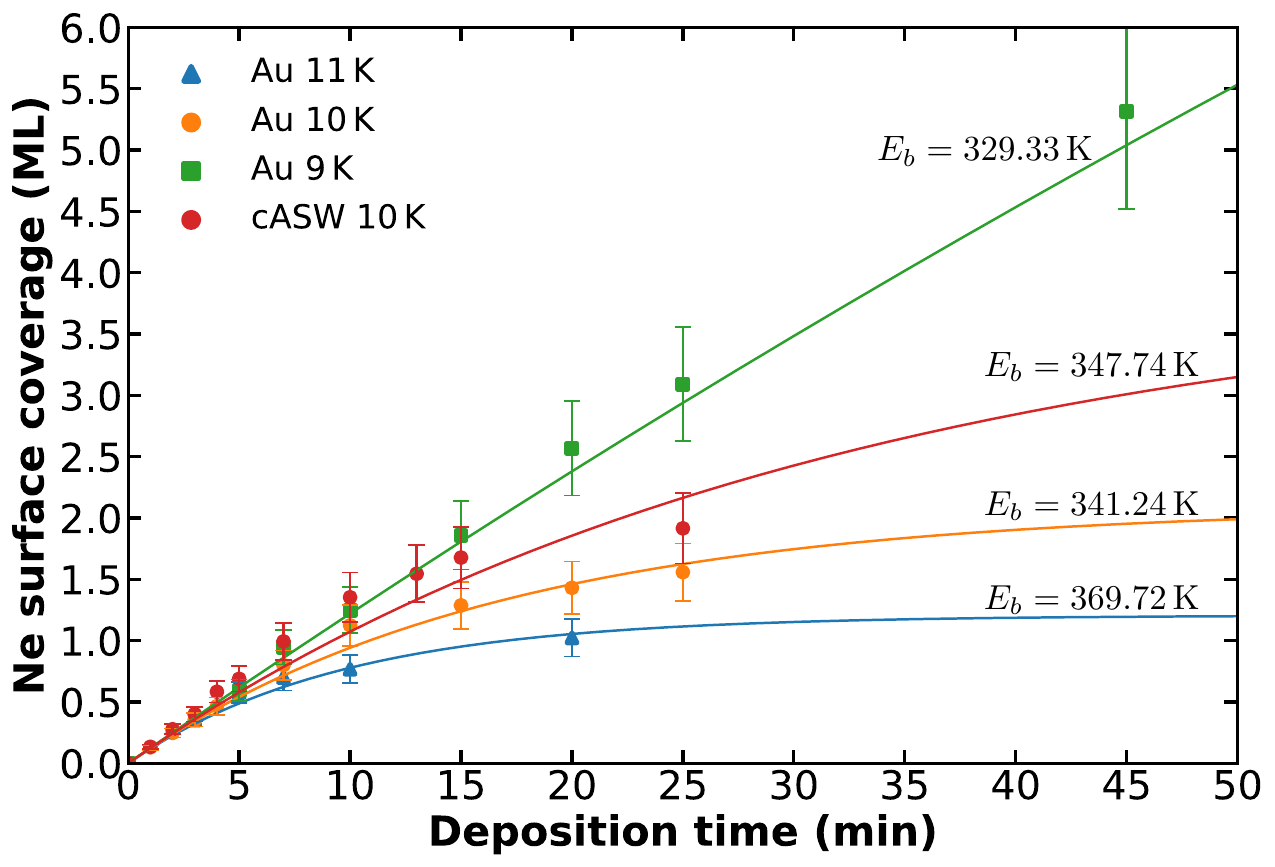}
    \caption{Evolution of the surface coverage of a layer of Ne according to the deposition time for different temperatures and substrates.}
    \label{figNe: saturation Neon}
\end{figure}

Figure\,\ref{figNe: saturation Neon} illustrates the surface coverage of the neon that remains on the surface as a function of the deposition time for different temperatures. Each point is represented with its error bar and corresponds to the surface coverage of Ne obtained with TPD after a given deposition time (from 1 to 50\,minutes) for different surfaces (Au or cASW) at different constant temperatures (9, 10, or 11\,K).
We observe a saturation regime caused by the increase in the desorption rate with the surface coverage that progressively compensates for the accretion rate. The plateau is reached when the desorption rate equals the accretion rate, at the end or during the deposition of Ne. At 11\,K on gold, neon only partially forms a monolayer, which can only be achieved at 10\,K, while several layers can be obtained at 9\,K. 
The solid line represents a simple rate equation model that considers a constant accretion and desorption rate given by
\begin{equation}
    dN_s/dt = \Phi_{\ce{Ne}} - N_s(t)Ae^{-E_b/T_s} \;,
\end{equation}
where $\Phi_{\ce{Ne}}$ represents the constant flux of Ne, $N_s(t)$ the surface coverage of Ne, $A$ the pre-exponential factor ($A=10^{12}$\,s$^{-1}$ as derived from \citealt{schlichting1993techniques}), $E_b$ the binding energy of Ne, and $T_s$ the surface temperature during the deposition. The experimental data were fitted by treating the binding energy as a free parameter. We found the best fit of the experiments using an effective binding energy of 369.72\,K (blue curve) for the case of an Au surface held at 11\,K, to approximately 330\,K for the 9\,K case (represented in green). In the first case (11\,K, blue), this only means that the surface coverage achieved at the end of the experiment (around one monolayer)  had an effective binding energy that cannot be lower than this value. This simple approach shows that the colder the surface, the higher the coverage and the lower the effective binding energy. The lower boundaries of the binding energy distribution lie in the range of 330-370\,K. In Appendix\,B, we analyse the binding energy of Ne in more detail.  

When we compare these results with the previous experiments (Figure\,\ref{figNe: TPD H2CO}), we note that the greater the quantity of neon on the surface, the smaller the surface coverage of \ce{H2CO} that formed on the surface. Neon has disturbed hydrogen atoms by blocking them as they diffuse across the surface. Consequently, the probability of hydrogen atoms encountering a carbon monoxide molecule is significantly reduced. On cASW, the binding energy of neon is slightly higher, as shown in Fig.\,\ref{figNe: saturation Neon}, where the quantity of neon saturates slightly above that on the gold substrate for the same temperature. Even though cASW has catalytic properties and/or retention capabilities that lead to a larger production of \ce{H2CO} and \ce{CH3OH} when there is no neon, the same chemical inhibition phenomenon is observed when Ne is co-deposited. The larger coverage of neon at a given temperature means, however, that we might expect a stronger inhibition, which is the case in absolute values (the reduction is larger), but not in relative values, as shown in Table\,\ref{tab: Rapport mesure}.

\begin{table}
    \caption{\ce{H2CO} and \ce{CH3OH} formation ratio ($N^{\{CO+H+Ne\}}_X$/$N^{\{CO+H\}}_X$) during \{CO+H+Ne\} co-deposition compared to \{CO+H\} co-deposition.}
    \label{tab: Rapport mesure}
    \centering
    \begin{tabular}{l c c  c c }
        \hline 
        & \multicolumn{2}{c}{\ce{H2CO}} & \multicolumn{2}{c}{\ce{CH3OH}} \\
        \hline
        T(K) & Au & cASW & Au & cASW \\
        \hline
        9 & 0.09 & 0.10 & 0.23 & 0.21  \\
        10 & 0.23 & 0.33 & 0.39 & 0.47 \\
        11 &  0.62 & & 0.57 & \\
        12 & 0.85 & & 1.02 & \\
        \hline
    \end{tabular}
\end{table} 

\begin{figure}
    \centering
    \includegraphics[width=\columnwidth]{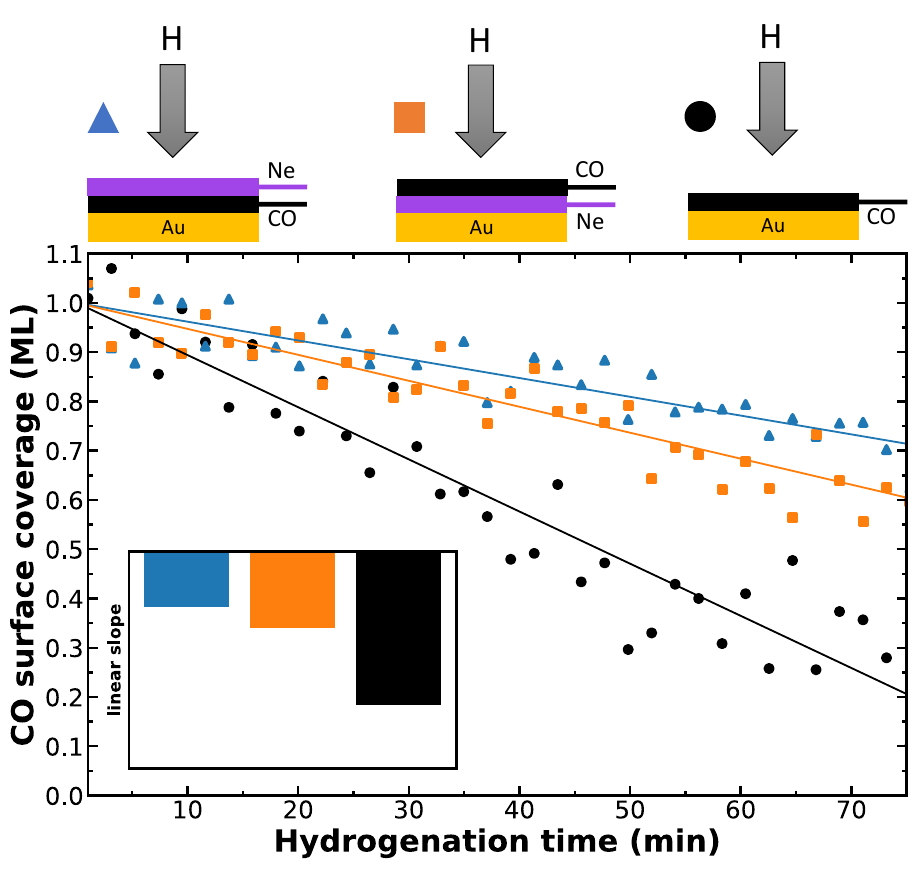}
    \caption{Integrated infrared-band area of CO in the case of H atom bombardment at 9\,K of pure CO ice (1\,ML, black circles), CO ice deposited onto Ne (1\,ML, 1.43\,ML, orange squares), and CO ice deposited under Ne (1.43\,ML, 1\,ML, blue triangles). The inset shows the linear slope of the fit from each experiment.}
    \label{figNe: CO decroissance}
\end{figure}

The quantity of molecules present on the surface can be retrieved from TPD data, but it can only be acquired after a deposition ends (see section\,\ref{sec: Experimental method}). To gain a deeper understanding of the impact of Ne on reactivity of CO molecules with H atoms, FT-RAIRS spectra can be generated because they are measured directly during the deposition process at two-minute intervals. The CO molecules are highly sensitive to infrared radiation and can therefore be readily quantified. Figure \ref{figNe: CO decroissance} illustrates the decay of a CO monolayer following atomic hydrogen bombardment over time for three distinct configurations: a CO monolayer below a Ne monolayer (blue triangles), a CO monolayer above a Ne monolayer (orange rectangles), and a CO monolayer alone (black circles). When Ne is situated above CO, and because it is not reactive with atomic hydrogen, only a small proportion of the CO reacts with the hydrogen, resulting in a gradual decrease. In the absence of Ne, the H atoms are free to react with CO molecules, which leads to a rapid decrease in the surface coverage of CO. 

\begin{figure}
    \centering
    \includegraphics[width=\columnwidth]{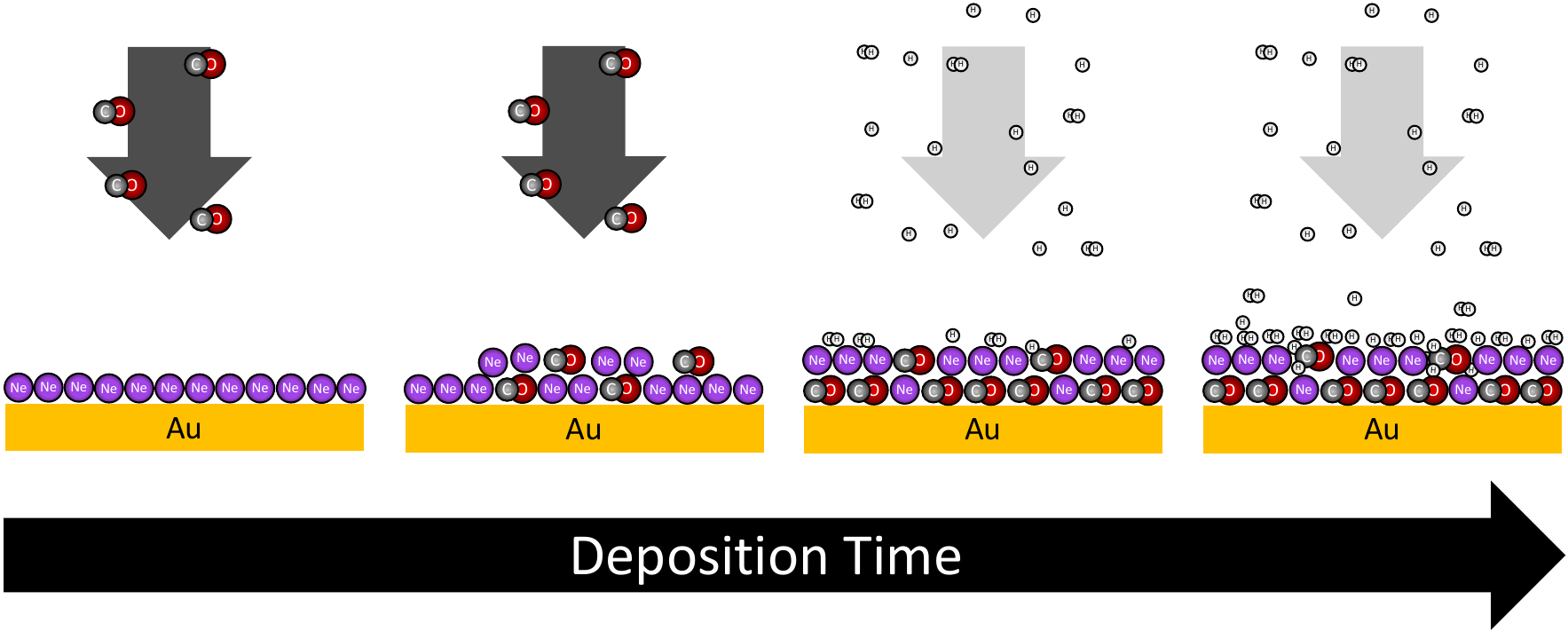}
    \caption{Schematic view of the physical process of Ne blocking hydrogenation as represented in Figure\,\ref{figNe: CO decroissance}. When CO molecules accrete onto the surface, they diffuse below the neon layer. In contrast to this, neon acts to prevent hydrogen from diffusing into the bulk, thus retaining it on the surface. This means that the H atom cannot react with CO.}
    \label{figNe: Schema neon}
\end{figure}

The most intriguing result is that of the case when Ne is deposited below CO. \cite{congiu2020new} demonstrated that in a layered experiment, only the molecules in the topmost layer of the surface interact with hydrogen atoms. Consequently, when CO is deposited on Ne, its decay due to hydrogen bombardment is expected to resemble that observed in the absence of Ne. Figure \ref{figNe: CO decroissance} shows, however, that the CO decay is slower by more than a factor of two and closer to the decay measured when Ne is above CO. Since the coverage of deposited ice does not exceed two to three monolayers, only two possible configurations exist for a Ne atom: It resides above a CO molecule, or beneath it. Therefore, we assume that when neon is deposited before CO, the two species form a mixture where the neon atoms tend to diffuse towards the ice surface, forming an impermeable layer to hydrogen atoms. A hydrogen atom reaches the Ne layer, diffuses across it, and recombines with another H, without reacting with the CO underneath. Our hypothesis is illustrated schematically in Figure\,\ref{figNe: Schema neon}. The situation is overly simplified, and there is a whole range of conditions that make hydrogen less accessible to \ce{CO} (orientation, sides, and steps). Figure\,\ref{figC: figure_detailes_fig4_5} in the appendix illustrates this continuum of situations.

\section{Astrophysical implications} \label{sec: Astrophysical implications}

\begin{figure}
    \centering
    \includegraphics[width=\columnwidth]{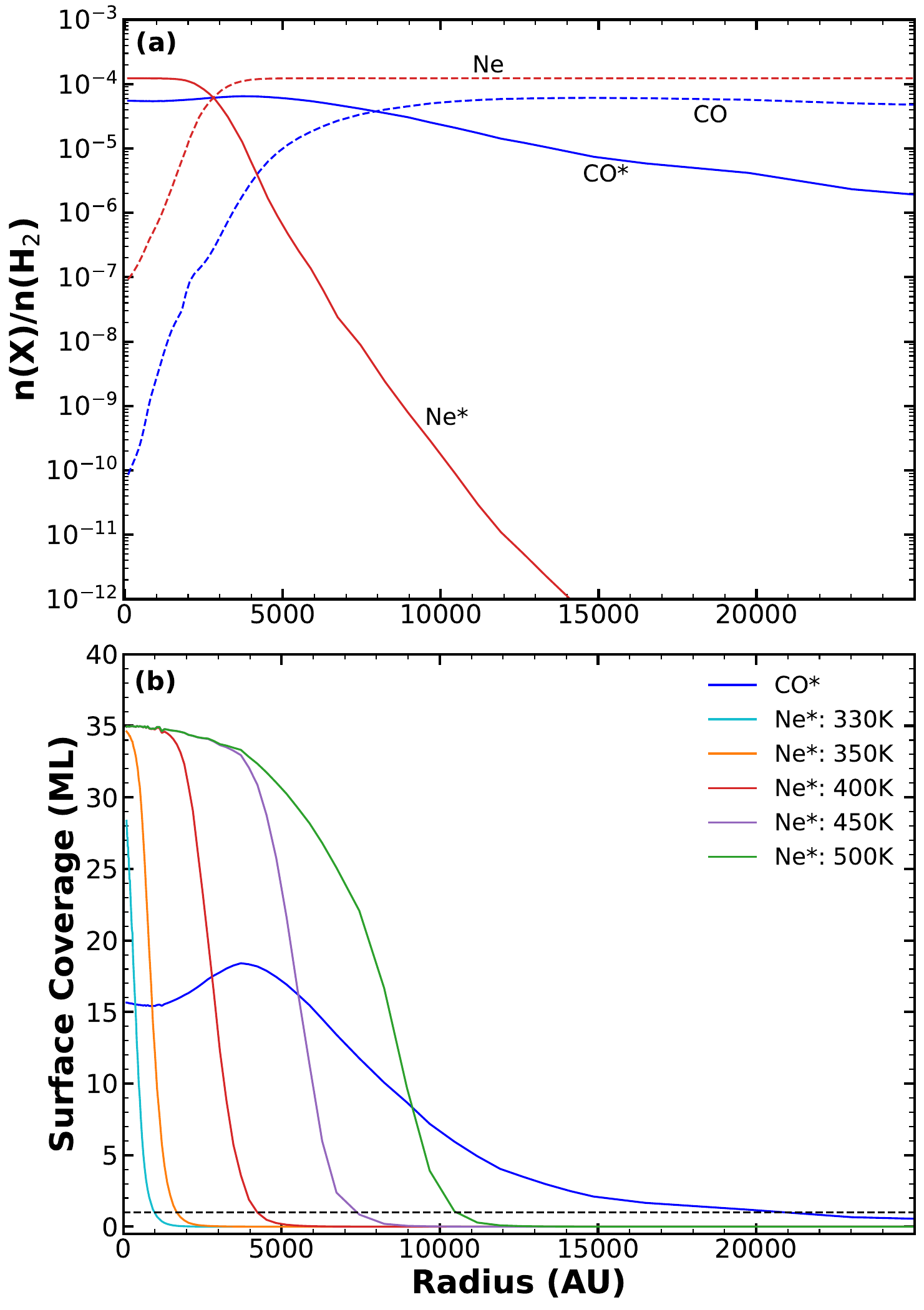}
    \caption{\textit{Upper panel}: Abundance of CO and Ne on the grain surface (solid line) and in the gas phase (dashed line) as a function of the distance from the centre of a pre-stellar core. The asterisks denote ice species. \textit{Lower panel}: Similar to the \textit{upper panel}, but focusing on the grain surface coverage for Ne with varying binding energy values, as indicated by the legend.} 
    \label{figNe: abn molecularCloud pyRate}
\end{figure}

We carried out a two-phase (gas and solid) chemical modelling of Ne to identify its condensation region in a pre-stellar core and estimate the surface coverage. The gas-grain chemical code pyRate, described by \cite{sipila2015benchmarking}, is based on rate equations for calculating the time evolution of chemical species in the gas and solid phase. The simulation used the physical model of a pre-stellar core presented by \cite{keto2014chemistry}, and the set of initial abundances were taken from \cite{semenov2010chemistry}. We chose a time of $t=2.5\times10^5$\,yr to align with the best fit to observational findings by \cite{caselli2022central}.

We chose an initial Ne abundance of $1.23\times10^{-4}$ ($n(\ce{Ne})/n(\ce{H})$) based on observational measurements collected by \cite{nicholls2017abundance} in our Milky Way and nearby dwarf galaxies. Neon does not freeze out until the dust temperature drops below 11\,K; thus, its presence is expected to only affect dense clouds, where the temperature is 10\,K or lower. 
As neon is a non-reactive element, the variation of its initial abundance only affects the final concentration in ice. Most of the gaseous Ne will simply accrete onto dust grains without reacting with other species. The upper panel of Figure\,\ref{figNe: abn molecularCloud pyRate} shows the gas and solid phase evolution of Ne and CO in a pre-stellar core (assuming a value of the binding energy of 400\,K for Ne and 1150\,K for CO).

The critical role of neon lies not in its final abundance on the grains, however, but in the completion of the first layer. Achieving this coverage greatly reduces the efficiency of hydrogenation processes, as demonstrated by our experiments with CO. The accretion rate is governed by the density of the gas and is balanced by the temperature of grains that governs the desorption (and to a lesser degree, by non-thermal desorption processes such as cosmic-ray desorption). The density increases while the temperature decreases towards the centre of the core, and both physical constraints lead to an increase in the efficiency of gas depletion towards the centre. 
Therefore, there is a radius within which the first layer of neon may be completed, but this is dependent on its binding energy. The lower panel of Figure \ref{figNe: abn molecularCloud pyRate} illustrates the surface coverage as the radial profile for the molecules of CO (blue) and Ne for specific binding energy values: 330\,K (cyan), 350\,K (orange), 400\,K (red), 450\,K (purple), and 500\,K (green). The binding energy for CO is set at $E_b=1150$\,K \citep{Noble2012}, and the dashed horizontal line represents one monolayer. Therefore, the intersection of this dashed line with the coloured lines corresponds to the radius inside which the hydrogenation is reduced. The choice of neon binding energy clearly affects the area where the neon can prevent hydrogenation. For a binding energy $E_b=330$\,K, the monolayer is reached at $R\sim1000$\,AU. In contrast, for $E_b=500$\,K, the monolayer is reached at $R\sim11000$\,AU with an intermediate radius of $R\sim5000$\,AU for $E_b=400$\,K. 
For most of the various binding energies assumed for Ne, an asymptote at 35\,ML is reached in the central part of the pre-stellar core, corresponding to the total condensation of gas-phase neon on grains. The value of the asymptote depends on the initial abundance of neon and the physical parameters of the grains (e.g. size and composition)

Neon features a range of binding energies onto interstellar ices (see Appendix \ref{app: Binding energy calculation} for the details), with an average of 425\,K. By interpreting the lower panel of Figure\,\ref{figNe: abn molecularCloud pyRate} for a model that includes a binding energy distribution, we found that neon starts to condense at radii near $R\sim11000$\,AU, achieves a monolayer coverage at $R\sim5000$\,AU, and additional layers develop at smaller radii, around $R\sim1000$\,AU. Consequently, we recommend setting the binding energy of Ne between 400\,K and 425\,K in current rate-equation models. Crucially, when the first layer is formed and the surface is saturated with neon, the chemistry is deeply modified. As a consequence, having either one monolayer or three monolayers of neon ice does not affect the final results much. We cannot currently evaluate the impact of the Ne freeze-out with the chemical model, however. To do this, we would need to modify our description of the surface layer and allow the neon to stay on top of the grain and not be included in the ice mantle, as we hypothesized in our Figure \ref{figNe: Schema neon}. This will be developed in a future project. 

Other non-reactive species in addition to neon might theoretically play a major role in modifying the effectiveness of hydrogenation onto dust grains. For this to be possible, the abundance of the species must be high enough for it to cover at least one monolayer on the grain. We can therefore estimate that given the effective surface area of the available grains, this species must have a relative abundance greater than $4\times10^{-6}$ \citep{cazaux2016}. Consequently, this excludes Ar and other noble gases, except He. The binding energy of the considered species must also be low enough for them to remain on the surface of the grains (and not in the mantle), however, but high enough not to desorb. This rules out He, whose binding energy is too low ($E_b=100$\,K). 

On the other hand, the H$_2$ molecule is very abundant, chemically inert, and has a range of binding energy distributions \citep{Amiaud2006}. Our experiments showed that \ce{H2} does not interfere with hydrogenation. Because our hydrogen atom beam is not create a perfectly dissociated plasma, we achieved a co-deposition of H and \ce{H2}, instead of H alone. Furthermore, in the absence of other reactants, H atoms react to form \ce{H2} at all temperatures, and a part of the molecules will remain on the surface. In the same way as Ne, \ce{H2} accumulates over time until it saturates the surface (i.e. when accretion equals desorption), although according to \cite{Amiaud2007}, it should saturate around a fraction of a monolayer ($\sim0.3$) at 10\,K. Nevertheless, we found that hydrogenation is not prevented at all at low temperatures; on the contrary, as shown in Figure \ref{figNe: TPD H2CO}, it is more efficient at the lowest temperatures as a result of the tunnelling effect, which is proof that \ce{H2}, although more abundant on the surface, does not hinder hydrogenation. Therefore, the reduction of the hydrogenation effectiveness we observed experimentally is probably specific to the neon in astrophysical environments such as dense cores.

\section{Conclusion} \label{sec: Conclusion}

We have experimentally shown that neon (Ne) can condense onto the surface of dust grains at a very low temperature (<12\,K) and significantly reduces the reactivity between hydrogen atoms and carbon monoxide on gold and compact amorphous solid-water substrates under pre-stellar core conditions. 
We found that the formation of formaldehyde (\ce{H2CO}) and methanol (\ce{CH3OH}) through CO hydrogenation can be reduced by as much as 90\,\%. This reduction is intrinsically connected to the surface coverage of Ne that is present on the surface. The effect on the reaction product formation becomes particularly significant when a neon monolayer is reached. 
We showed that regardless of the order in which Ne and CO are deposited, the formation of \ce{H2CO} and \ce{CH3OH} is significantly diminished. We hypothesized that Ne can diffuse onto the grains and preferentially accumulates on the surface. This behaviour likely creates an insulating layer between the CO molecules and the upcoming H atoms. 
Finally, we used a gas-grain chemical code to show that Ne can condense in the central region (<5000\,AU) of pre-stellar cores, which means that the inhibiting effect of Ne can be very significant deep inside dense cores.
In our upcoming study, we will integrate the physical neon process into our gas-grain chemical code and analyse its effects on chemistry, with a focus on the formation of complex organic molecule in pre-stellar core models.

\begin{acknowledgements} 
    We thank our colleagues Valeriia Sokolova, Saoud Baouche, and Zakaria Idrissi-Machichi for their help with the experiments conducted for this work, and Stephan Diana for his support in IT.
    This work was supported by the Agence Nationale de la Recherche (ANR) SIRC project (Grant ANR-SPV202448 2020-2024). 
\end{acknowledgements}

\bibliographystyle{aa}
\bibliography{bibli.bib} 

\begin{thebibliography}{46}
\expandafter\ifx\csname natexlab\endcsname\relax\def\natexlab#1{#1}\fi

\bibitem[{Amiaud {et~al.}(2007)Amiaud, Dulieu, Fillion, Momeni, \&
  Lemaire}]{Amiaud2007}
Amiaud, L., Dulieu, F., Fillion, J.-H., Momeni, A., \& Lemaire, J.~L. 2007, J.
  Chem. Phys., 127, 144709

\bibitem[{Amiaud {et~al.}(2006)Amiaud, Fillion, Baouche, Dulieu, Momeni, \&
  Lemaire}]{Amiaud2006}
Amiaud, L., Fillion, J.~H., Baouche, S., {et~al.} 2006, J. Chem. Phys., 124,
  94702

\bibitem[{Armah {et~al.}(2021)Armah, Dors, Aydar, Cardaci, H{\"a}gele, Feltre,
  Riffel, Riffel, \& Krabbe}]{armah2021chemical}
Armah, M., Dors, O.~L., Aydar, C., {et~al.} 2021, MNRAS, 508, 371

\bibitem[{Bacmann {et~al.}(2002)Bacmann, Lefloch, Ceccarelli, Castets,
  Steinacker, \& Loinard}]{bacmann2002degree}
Bacmann, A., Lefloch, B., Ceccarelli, C., {et~al.} 2002, A\&A, 389, L6

\bibitem[{Bergin \& Tafalla(2007)}]{bergin2007cold}
Bergin, E.~A. \& Tafalla, M. 2007, ARA\&A, 45, 339

\bibitem[{Boogert {et~al.}(2015)Boogert, Gerakines, \&
  Whittet}]{boogert2015observations}
Boogert, A.~A., Gerakines, P.~A., \& Whittet, D.~C. 2015, ARA\&A, 53, 541

\bibitem[{Brinchmann {et~al.}(2013)Brinchmann, Charlot, Kauffmann, Heckman,
  White, \& Tremonti}]{brinchmann2013estimating}
Brinchmann, J., Charlot, S., Kauffmann, G., {et~al.} 2013, MNRAS, 432, 2112

\bibitem[{Butscher {et~al.}(2015)Butscher, Duvernay, Theule, Danger, Carissan,
  Hagebaum-Reignier, \& Chiavassa}]{butscher2015formation}
Butscher, T., Duvernay, F., Theule, P., {et~al.} 2015, MNRAS, 453, 1587

\bibitem[{Cameron(1973)}]{cameron1973abundances}
Cameron, A.~G. 1973, Space Sci. Rev., 15, 121

\bibitem[{Caselli {et~al.}(2022)Caselli, Pineda, Sipil{\"a}, Zhao, Redaelli,
  Spezzano, Maureira, Alves, Bizzocchi, Bourke, {et~al.}}]{caselli2022central}
Caselli, P., Pineda, J.~E., Sipil{\"a}, O., {et~al.} 2022, ApJ, 929, 13

\bibitem[{Caselli {et~al.}(1999)Caselli, Walmsley, Tafalla, Dore, \&
  Myers}]{caselli1999co}
Caselli, P., Walmsley, C., Tafalla, M., Dore, L., \& Myers, P. 1999, ApJ, 523,
  L165

\bibitem[{Cazaux {et~al.}(2016)Cazaux, Minissale, Dulieu, \&
  Hocuk}]{cazaux2016}
Cazaux, S., Minissale, M., Dulieu, F., \& Hocuk, S. 2016, A\&A, 585

\bibitem[{Cazaux \& Tielens(2002)}]{cazaux2002molecular}
Cazaux, S. \& Tielens, A. 2002, ApJ, 575, L29

\bibitem[{Chaabouni {et~al.}(2018)Chaabouni, Diana, Nguyen, \&
  Dulieu}]{chaabouni2018thermal}
Chaabouni, H., Diana, S., Nguyen, T., \& Dulieu, F. 2018, A\&A, 612, A47

\bibitem[{Congiu {et~al.}(2020)Congiu, Sow, Nguyen, Baouche, \&
  Dulieu}]{congiu2020new}
Congiu, E., Sow, A., Nguyen, T., Baouche, S., \& Dulieu, F. 2020, Rev. Sci.
  Instrum., 91

\bibitem[{Crapsi {et~al.}(2007)Crapsi, Caselli, Walmsley, \&
  Tafalla}]{crapsi2007observing}
Crapsi, A., Caselli, P., Walmsley, M.~C., \& Tafalla, M. 2007, A\&A, 470, 221

\bibitem[{Dors~Jr {et~al.}(2013)Dors~Jr, H{\"a}gele, Cardaci,
  P{\'e}rez-Montero, Krabbe, V{\'\i}lchez, Sales, Riffel, \&
  Riffel}]{dors2013optical}
Dors~Jr, O.~L., H{\"a}gele, G.~F., Cardaci, M.~V., {et~al.} 2013, MNRAS, 432,
  2512

\bibitem[{Dulieu {et~al.}(2019)Dulieu, Nguyen, Congiu, Baouche, \&
  Taquet}]{Dulieu2019formamide}
Dulieu, F., Nguyen, T., Congiu, E., Baouche, S., \& Taquet, V. 2019, MNRAS,
  484, L119

\bibitem[{Fedoseev {et~al.}(2015)Fedoseev, Cuppen, Ioppolo, Lamberts, \&
  Linnartz}]{fedoseev2015experimental}
Fedoseev, G., Cuppen, H.~M., Ioppolo, S., Lamberts, T., \& Linnartz, H. 2015,
  MNRAS, 448, 1288

\bibitem[{Gould \& Salpeter(1963)}]{gould1963interstellar}
Gould, R.~J. \& Salpeter, E.~E. 1963, ApJ, 138, 393

\bibitem[{Grieco {et~al.}(2023)Grieco, Theul{\'e}, De~Looze, \&
  Dulieu}]{grieco2023enhanced}
Grieco, F., Theul{\'e}, P., De~Looze, I., \& Dulieu, F. 2023, Nat. Astron., 7,
  541

\bibitem[{Herbst \& Van~Dishoeck(2009)}]{herbst2009complex}
Herbst, E. \& Van~Dishoeck, E.~F. 2009, ARA\&A, 47, 427

\bibitem[{Hollenbach \& Salpeter(1971)}]{hollenbach1971surface}
Hollenbach, D. \& Salpeter, E. 1971, ApJ, 163, 155

\bibitem[{Hwang {et~al.}(1996)Hwang, Kim, \& Rudd}]{hwang1996new}
Hwang, W., Kim, Y.-K., \& Rudd, M.~E. 1996, J. Chem. Phys., 104, 2956

\bibitem[{Ioppolo {et~al.}(2021)Ioppolo, Fedoseev, Chuang, Cuppen, Clements,
  Jin, Garrod, Qasim, Kofman, Van~Dishoeck, {et~al.}}]{ioppolo2021non}
Ioppolo, S., Fedoseev, G., Chuang, K.-J., {et~al.} 2021, Nat. Astron., 5, 197

\bibitem[{Keto \& Caselli(2008)}]{keto2008different}
Keto, E. \& Caselli, P. 2008, ApJ, 683, 238

\bibitem[{Keto {et~al.}(2014)Keto, Rawlings, \& Caselli}]{keto2014chemistry}
Keto, E., Rawlings, J., \& Caselli, P. 2014, MNRAS, 440, 2616

\bibitem[{Kruczkiewicz {et~al.}(2024)Kruczkiewicz, Dulieu, Ivlev, Caselli,
  Giuliano, Ceccarelli, \& Theul{\'e}}]{kruczkiewicz2024comprehensive}
Kruczkiewicz, F., Dulieu, F., Ivlev, A., {et~al.} 2024, A\&A, 686, A236

\bibitem[{Kumar {et~al.}(2019)Kumar, Kumar, Kumar, \&
  Kumar}]{kumar2019electron}
Kumar, Y., Kumar, M., Kumar, S., \& Kumar, R. 2019, Atoms, 7, 60

\bibitem[{Launhardt {et~al.}(2013)Launhardt, Stutz, Schmiedeke, Henning,
  Krause, Balog, Beuther, Birkmann, Hennemann, Kainulainen,
  {et~al.}}]{launhardt2013earliest}
Launhardt, R., Stutz, A., Schmiedeke, A., {et~al.} 2013, A\&A, 551, A98

\bibitem[{{Lodders}(2010)}]{lodders2010solar}
{Lodders}, K. 2010, in Astrophysics and Space Science Proceedings, Vol.~16,
  Principles and Perspectives in Cosmochemistry, 379

\bibitem[{Minissale {et~al.}(2022)Minissale, Aikawa, Bergin, Bertin, Brown,
  Cazaux, Charnley, Coutens, Cuppen, Guzman, {et~al.}}]{Minissale2022Review}
Minissale, M., Aikawa, Y., Bergin, E., {et~al.} 2022, ACS Earth Space Chem., 6,
  597

\bibitem[{Nicholls {et~al.}(2017)Nicholls, Sutherland, Dopita, Kewley, \&
  Groves}]{nicholls2017abundance}
Nicholls, D.~C., Sutherland, R.~S., Dopita, M.~A., Kewley, L.~J., \& Groves,
  B.~A. 2017, MNRAS, 466, 4403

\bibitem[{Noble {et~al.}(2012)Noble, Congiu, Dulieu, \& Fraser}]{Noble2012}
Noble, J.~A., Congiu, E., Dulieu, F., \& Fraser, H.~J. 2012, MNRAS, 421, 768

\bibitem[{Pagani {et~al.}(2007)Pagani, Bacmann, Cabrit, \&
  Vastel}]{pagani2007depletion}
Pagani, L., Bacmann, A., Cabrit, S., \& Vastel, C. 2007, A\&A, 467, 179

\bibitem[{Pagani {et~al.}(2012)Pagani, Bourgoin, \& Lique}]{pagani2012method}
Pagani, L., Bourgoin, A., \& Lique, F. 2012, A\&A, 548, L4

\bibitem[{Rejoub {et~al.}(2002)Rejoub, Lindsay, \&
  Stebbings}]{rejoub2002determination}
Rejoub, R., Lindsay, B., \& Stebbings, R. 2002, Phys. Rev. A, 65, 042713

\bibitem[{Schlichting \& Menzel(1993)}]{schlichting1993techniques}
Schlichting, H. \& Menzel, D. 1993, Rev. Sci. Instrum., 64, 2013

\bibitem[{Semenov {et~al.}(2010)Semenov, Hersant, Wakelam, Dutrey, Chapillon,
  Henning, Launhardt, Pi{\'e}tu, Schreyer, {et~al.}}]{semenov2010chemistry}
Semenov, D., Hersant, F., Wakelam, V., {et~al.} 2010, A\&A, 522, A42

\bibitem[{Sipil{\"a} {et~al.}(2015)Sipil{\"a}, Caselli, \&
  Harju}]{sipila2015benchmarking}
Sipil{\"a}, O., Caselli, P., \& Harju, J. 2015, A\&A, 578, A55

\bibitem[{Sofia {et~al.}(1994)Sofia, Cardelli, \& Savage}]{sofia1994abundant}
Sofia, U.~J., Cardelli, J.~A., \& Savage, B.~D. 1994, ApJ, 430, 650

\bibitem[{Tielens \& Hagen(1982)}]{tielens1982model}
Tielens, A. \& Hagen, W. 1982, \aap, 114, 245

\bibitem[{Tsuge \& Watanabe(2023)}]{tsuge2023radical}
Tsuge, M. \& Watanabe, N. 2023, Proc. Jpn. Acad. Ser. B, 99, 103

\bibitem[{Vinodkumar {et~al.}(2011)Vinodkumar, Bhutadia, Limbachiya, \&
  Joshipura}]{vinodkumar2011electron}
Vinodkumar, M., Bhutadia, H., Limbachiya, C., \& Joshipura, K. 2011, Int. J.
  Mass Spectrom., 308, 35

\bibitem[{Vitorino {et~al.}(2024)Vitorino, Loison, Wakelam, Congiu, \&
  Dulieu}]{vitorino2024sulphur}
Vitorino, J., Loison, J.-C., Wakelam, V., Congiu, E., \& Dulieu, F. 2024,
  MNRAS, 533, 52

\bibitem[{Wakelam {et~al.}(2017)Wakelam, Bron, Cazaux, Dulieu, Gry, Guillard,
  Habart, Hornek{\ae}r, Morisset, Nyman, {et~al.}}]{wakelam2017h2}
Wakelam, V., Bron, E., Cazaux, S., {et~al.} 2017, Mol. Astrophys., 9, 1

\end{thebibliography}

\begin{appendix}

\section{Experimental set-up} \label{app: Experimental set-up}

\begin{figure*}
    \centering
    \includegraphics[width=\textwidth]{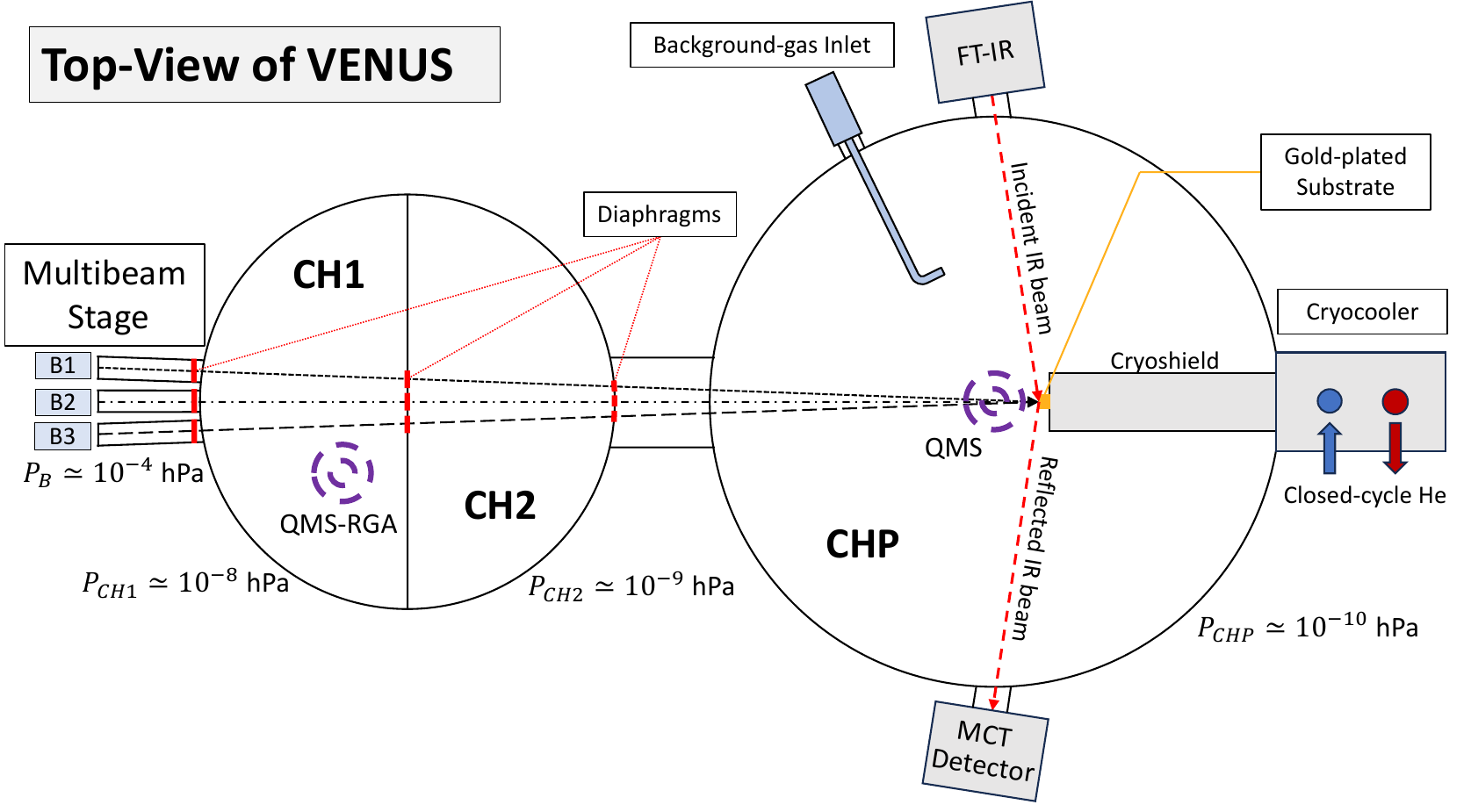}
    \caption{Schematic top-view of VENUS. The apparatus includes three gas beams (B1, B2, and B3), collimated on the gold surface of the main chamber by three series of diaphragms. A quadrupole mass spectrometer with a residual gas analyser (QMS-RGA) is situated within the chamber n°1 (CH1) to monitor the stability, composition, and purity of the gas flows during the course of an experiment. A second deposition system is connected to the main chamber (CHP) to enable background vapor deposition. The pressures in CH1 and CH2 are $10^{-8}$ and $10^{-9}$\,hPa, respectively, and the pressure in CHP reaches $10^{-10}$\,hPa. Atoms or molecules deposited on the surface can be quantified by the FT-RAIRS system during deposition and/or by the QMS in the main chamber, which is positioned in front of the surface, and a TPD can be performed.}
    \label{figA: Schematic VENUS}
\end{figure*}

The VENUS (VErs de NoUvelles Synthèses) apparatus is a multi-beam ultra-high vacuum (UHV) system located at LIRA-CY laboratory at CY Cergy Paris University. VENUS has been designed to study diffusion and surface reactivity to better understand the formation of complex organic molecules (COMs) through non-energetic processes in dark cloud conditions in the interstellar medium (ISM). A detailed technical description of VENUS is given in \cite{congiu2020new}. A schematic description is provided in Figure\,\ref{figA: Schematic VENUS}, and the essential elements are described below.

VENUS is made up of three stainless steel chambers and five beam chambers, out of which three are used for this study. The intermediate chambers (CH1 and CH2) are under high-vacuum (HV) conditions, with pressures of $10^{-8}$\,hPa and $10^{-9}$\,hPa respectively. The main chamber (CHP) is under ultra-high-vacuum (UHV) conditions, with pressures approaching $10^{-10}$\,hPa. The beam chambers have a pressure of $10^{-5}-10^{-4}$\,hPa while operating, depending on the required flow.

The gases in the beam chambers are collimated onto the cold gold polycrystal target thanks to the pressure gradient in the different chambers and the presence of a series of three 2\,mm-diaphragms. For commonly used molecules such as \ce{CO}, \ce{NO} and \ce{H2O}, a pressure of $10^{-4}$\,hPa in the beam allows a monolayer to be formed on the substrate in about 10\,minutes. A Leybold 100S quadrupole mass spectrometer residual gas analyser (QMS-RGA) is positioned in chamber one (CH1) to measure any flow fluctuations or leaks in the beams. A leakage valve connected to the main chamber allows background gas injection to deposit molecules by changing the partial pressure in the main chamber. This deposition process is mainly used to form water substrates with different morphologies such as porous-ASW, compact-ASW, or crystalline ice (protocol described in \citealt{congiu2020new}).

The molecules are deposited on a sample holder, which consists of a gold-coated 9\,mm diameter copper mirror connected to the cold head of a closed-cycle helium cryostat. The surface temperature is measured by a silicon diode, ranging from 7\,K to 400\,K, and monitored by a Lakeshore 340 controller with a stability of $\pm0.1$\,K and an accuracy of $\pm1$\,K.

\begin{figure}
    \centering
    \includegraphics[width=\columnwidth]{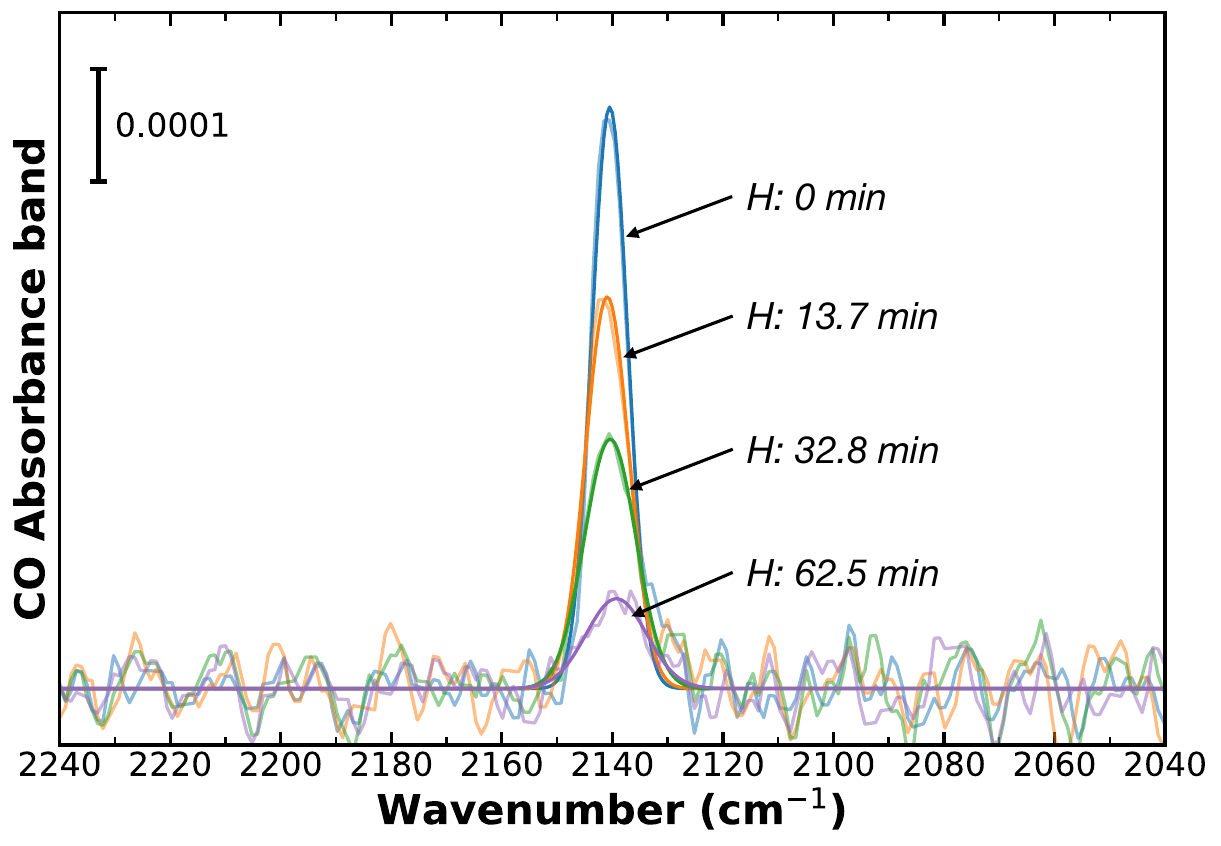}
    \caption{Various RAIRS spectra of CO band (2140\,cm$^{-1}$) measured for different hydrogenation times.}
    \label{figA: CO IRband}
\end{figure}

During the deposition, the evolution of the ice can be monitored using a Vertex-70 spectrometer equipped with Fourier Transform Reflected Absorption Infrared Spectroscopy (FT-RAIRS). The spectrometer laser is positioned at an angle of 83° to the surface, which enhances its sensitivity by a factor of five compared to transmission spectroscopy. However, determining the optical constants with this system is extremely challenging. Figure \ref{figA: CO IRband} illustrates the spectral evolution of a 1\,ML-thick layer of CO ice bombarded by hydrogen atoms; the area evolution is represented by the points and black curve in Figure \ref{figNe: CO decroissance}. The area of the CO band is determined by fitting each spectrum with a Gaussian. Infrared calibration is conducted using the Hiden 51/3F quadrupole mass spectrometer (QMS).

The QMS is a spectrometer that measures a molecule mass-to-charge ratio ($m/z$) in the gas phase in the VENUS main chamber and can be vertically moved. It is mainly used in two configurations. Firstly, in the low position, it measures the residual gas in the chamber. Secondly, in the high position, the QMS is placed between the beams and the surface (5\,mm from the latter), enabling the measurement of the beam flow before an experiment, or the molecules coming out of the surface during a temperature-programmed desorption (TPD, usually the last phase of an experiment). The latter is a method consisting in measuring the quantity of molecules desorbing from the surface while its temperature is linearly increased. Molecules can be identified by their desorption temperature as each has a different binding energy with the surface, as well as their mass cracking pattern. To minimize molecule fragmentation, we use an electron energy impact of 30\,eV in the QMS.

\section{Binding energy calculation} \label{app: Binding energy calculation}

\begin{figure}
    \centering
    \includegraphics[width=\columnwidth]{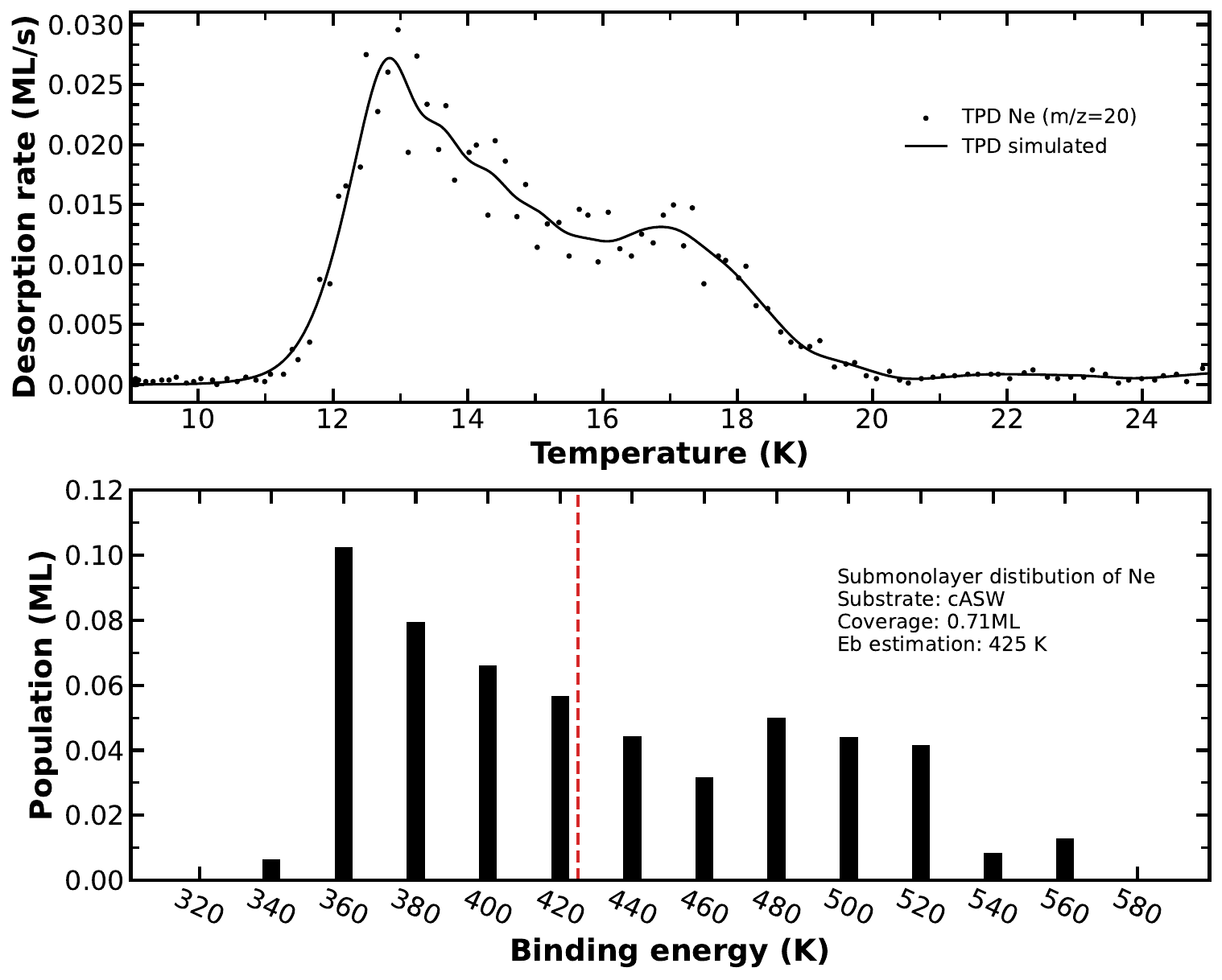}
    \caption{Distribution of Ne on a cASW substrate with a surface temperature at 9\,K during the deposition. The top panel represents a TPD spectrum of Ne ($m/z$=20) in black dots and a TPD simulated in black line. The bottom panel represents the distribution of binding energies of Ne on the cASW substrate. }
    \label{figB: Eb_submonolayer_SH2O}
\end{figure}

The Temperature Programmed Desorption (TPD) curve illustrates the distribution of adatoms or molecules on the surface. When a molecule is adsorbed onto the surface, it can diffuse to the sites with the highest binding energy, if not during the deposition phase, at least during the TPD phase. Consequently, the molecules are distributed across a range of energy sites, from the most bound to the least bound. 
The analysis of TPD data to calculate the distribution of binding energies was conducted using a custom-made software developed at LIRA-CY by S. Diana, following the model described by \cite{chaabouni2018thermal}. Briefly, the thermal desorption process is defined by the Polanyi-Wigner equation:
\begin{equation}
    k(T) = -\frac{dN}{dt} = AN^ae^{-E/k_bT}\,,
    \label{eq: Polanyi-Wigner}
\end{equation}
where $A$ is the exponential pre-factor, $N$ the surface coverage of molecules on the solid phase, $a$ the order of the desorption process, $E$ the binding energy, $k_b$ the Boltzmann constant and $T$ the grain temperature. The order of the desorption process, $a$, determines the regime of desorption. If $a = 0$, the desorption does not depend on the coverage, corresponding to a multilayer ice. When $a = 1$, the coverage is considered in equation \ref{eq: Polanyi-Wigner}, which represents an interaction between the surface and the molecules in a sub-monolayer regime \citep{Minissale2022Review}. In the following, we assume first-order desorption ($a = 1$), which will be transformed in the 0th order when the monolayer is reached.
Since adsorption sites do not all have the same energy, the coverage $N$ can be defined as the sum of different fractions of coverage $N = \sum^n_{i=1}N_i=N_1+N_2+...+N_n$. Each fraction of coverage $N_i$ has a specific binding energy $E_i$. The equation \ref{eq: Polanyi-Wigner} can be rewritten as: 
\begin{equation}
    k(T) = \sum^n_{i=1}k_i(T) = AN_1e^{-E_1/k_bT} + AN_2e^{-E_2/k_bT} + ... +AN_ne^{-E_n/k_bT}\,,
\end{equation}
where $k_i(T)$ is the desorption rate of the molecular population. The top panel of figure\,\ref{figB: Eb_submonolayer_SH2O} shows the best fit of a TPD curve of $0.71$\,ML of neon on a water ice analogue, compact amorphous solid water substrate (cASW) with a surface temperature held at 9\,K. The best fit is obtained using a pre-exponential factor of $A=10^{12}$\,s$^{-1}$ derived from \cite{schlichting1993techniques}. The fraction of coverage $N_i$ with their associated binding energy $E_i$ are represented in figure\,\ref{figB: Eb_submonolayer_SH2O} in the bottom panel.  
The binding energy of neon ranges from about 340\,K to 560\,K, with the lowest energy indicating the weakest binding site where neon can stick in the monolayer. The mean binding energy, associated with the multilayer, was estimated by \cite{schlichting1993techniques} to be around 267\,K. To obtain this binding energy, it would require a surface temperature close to 6\,K, which is not possible on VENUS. The coverage in figure\,\ref{figB: Eb_submonolayer_SH2O} corresponds to 0.71\,ML, indicating that the monolayer is not saturated, and consequently, the site with an energy of 340\,K is only partially occupied. This partial occupancy impacts the estimated average binding energy at 425\,K, as shown in the figure, resulting in a slight overestimation for the monolayer. Therefore, the mean binding energy for the monolayer is estimated to lie approximately between 400 and 425\,K.

\newpage
\section{Additional figures}

\begin{figure}
    \centering
    \includegraphics[width=\columnwidth]{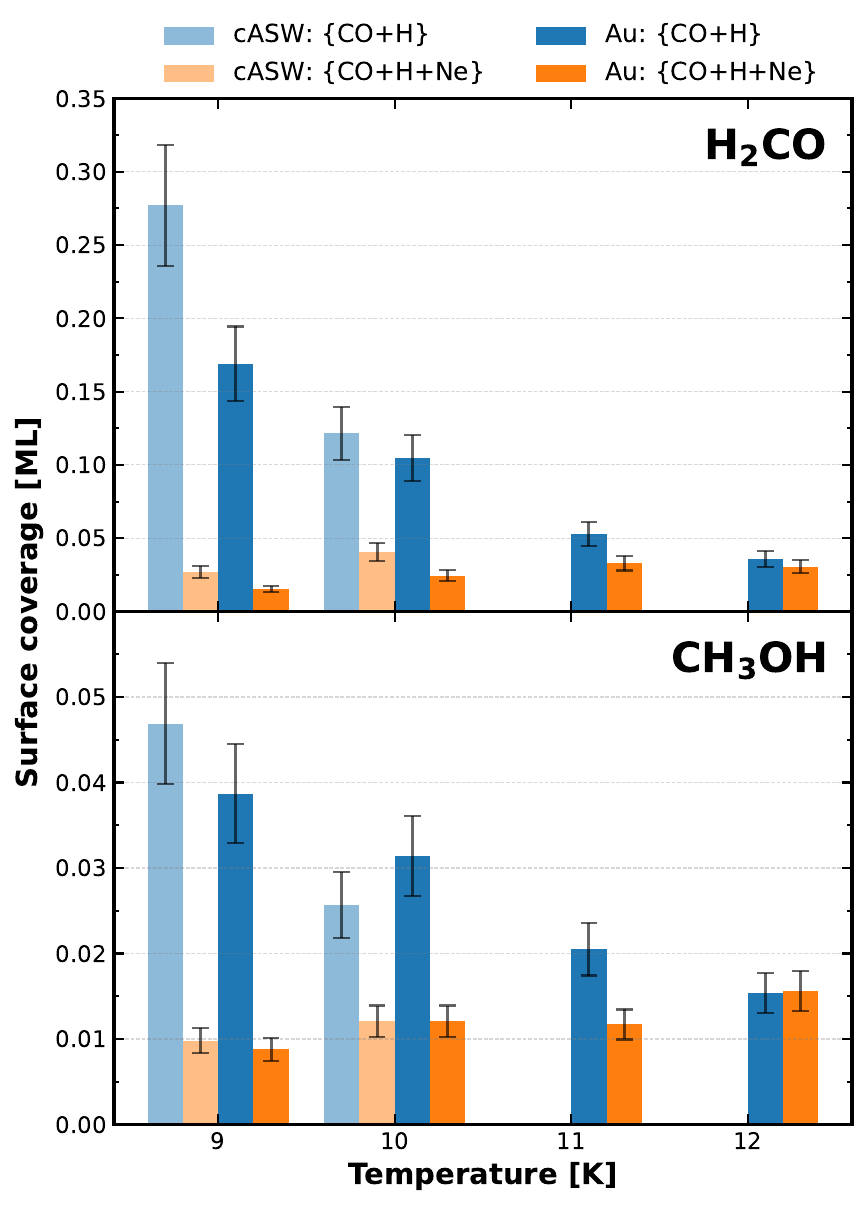}
    \caption{Surface coverage of \ce{H2CO} (top panel) and \ce{CH3OH} (bottom panel) across four different surface temperatures: 9, 10, 11 and 12\,K. Light blue and light orange bars represent experimental results of \ce{CO} co-deposition with \ce{H} (noted as \{\ce{CO + H}\}) and with \ce{H} and \ce{Ne} (as \{\ce{CO + H + Ne}\}) on a cASW substrate, respectively. Conversely, dark blue and dark orange bars correspond to the co-deposition on a gold (Au) substrate.}
    \label{figC: TPD_Yield_SAu_SASW}
\end{figure}

Figure\,\ref{figC: TPD_Yield_SAu_SASW} presents the surface coverage of \ce{H2CO} and \ce{CH3OH} during co-deposition of \{\ce{CO + H}\} and \{\ce{CO + H + Ne}\} at varying temperatures. These values were derived from the TPD curves illustrated in Figures\,\ref{figNe: TPD H2CO} and \ref{figNe: TPD CH3OH} for a gold substrate, using Equation\,\ref{eq: Nx}. The temperature and the properties of the substrate significantly impact the surface coverage of \ce{H2CO} and \ce{CH3OH}. These effects result from changes in ice conditions, such as sticking, diffusion, reactivity, or chemical desorption. For example, on Au substrate between $9$ and $12$\,K, \ce{H2CO} surface coverage decreases by about a factor of $4.7$, from $0.169$ to $0.036$\,ML. Similarly, \ce{CH3OH} surface coverage drops by approximately $2.5$ times, from $0.039$ to $0.015$\,ML. In the cASW substrate, the measured coverage of \ce{H2CO} is higher compared to that on the Au substrate. For \ce{CH3OH}, this effect is less pronounced, especially at $10$\,K, where the surface coverage on water is lower than on Au, but still remains in the experimental uncertainties. Despite these variations, the presence of neon on the surface has a fundamental impact on the reactivity of \ce{CO} with \ce{H}. At $9$\,K on Au, a $91$\% reduction in the formation of \ce{H2CO} and $77$\% of \ce{CH3OH} is measured (see Table\,\ref{tab: Rapport mesure}).

\begin{figure}
    \centering
    \includegraphics[width=\columnwidth]{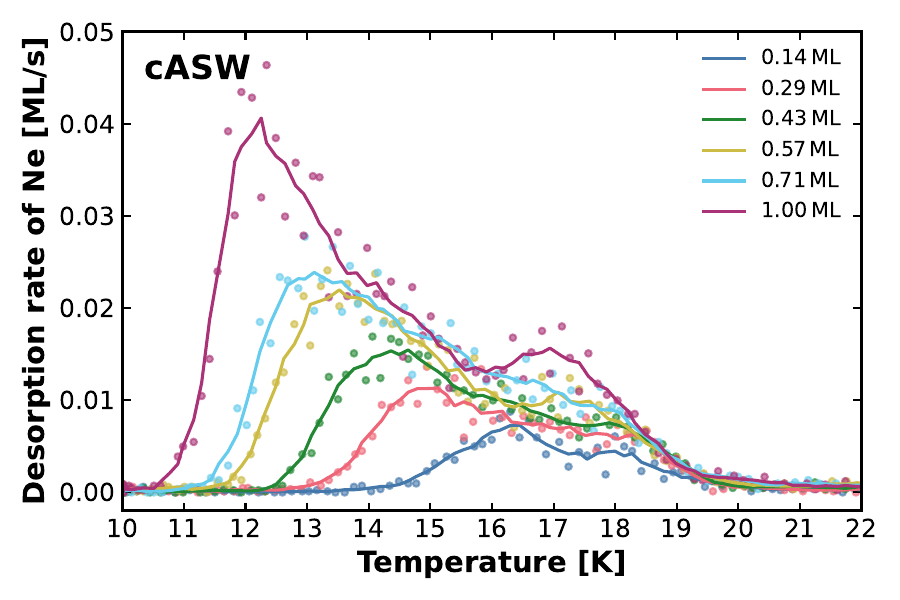}
    \caption{TPD curves of Ne ($m/z=20$) for varying surface coverages on cASW substrate at a constant surface deposition temperature of 10\,K. An error of 15\% is considered for the surface coverage.}
    \label{figC: figure_TPD_Ne_sublayer_H2O}
\end{figure}

\begin{figure}
    \centering
    \includegraphics[width=\columnwidth]{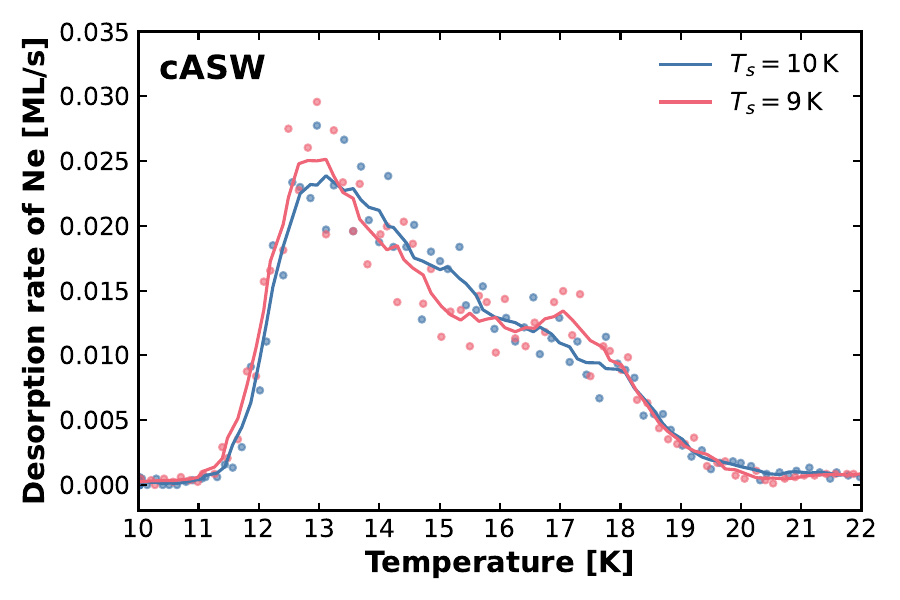}
    \caption{TPD curves of Ne ($m/z=20$) for a surface coverage of $0.71$\,ML on cASW substrate. The blue TPD curve represent a surface temperature during the deposition at 10\,K and 9\,K for the red TPD curve.}
    \label{figC: figure_TPD_Ne_sublayer_H2O_10K_9K}
\end{figure}

The Figure\,\ref{figC: figure_TPD_Ne_sublayer_H2O} shows the TPD curves of Ne for various coverages deposited on cASW at 10\,K. These curves correspond to the red point in Figure\,\ref{figNe: saturation Neon}, where the coverage is calculated using Equation\,\ref{eq: Nx}. Additionally, the TPD curves for 0.71\,ML of Ne deposited on cASW are presented in figure\,\ref{figC: figure_TPD_Ne_sublayer_H2O_10K_9K}, with the red curve representing a surface temperature during deposition of 9\,K and the blue curve corresponding to 10\,K. Considering the experimental uncertainty, no significant differences are observed between the two curves, indicating that no saturation of the Ne sublayer is detected at a surface temperature of 10\,K.

\begin{figure*}
    \centering
    \includegraphics[width=\textwidth]{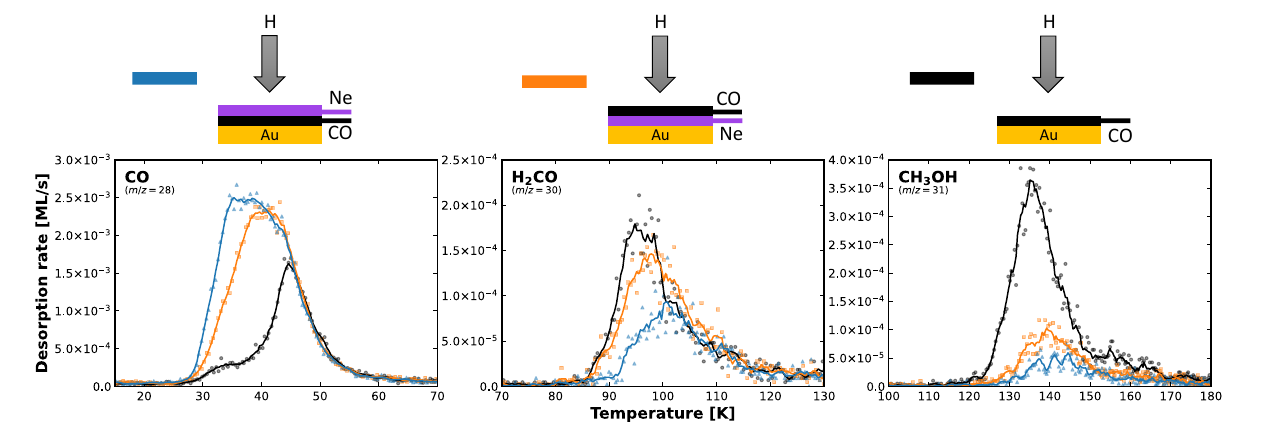}
    \caption{TPD curves of CO (left panel), \ce{H2CO} (middle panel) and \ce{CH3OH} (right panel) in the case of H atom bombardment at 9\,K of pure CO ice (1\,ML, black curves), CO ice deposited onto Ne (1\,ML, 1.43\,ML, blue curves), CO ice deposited under Ne (1.43\,ML, 1\,ML, orange curves).}
    \label{figC: figure_TPD_layerExp_legends}
\end{figure*}

Figure\,\ref{figC: figure_TPD_layerExp_legends} shows the TPD curves of CO (left panel), \ce{H2CO} (middle panel), and \ce{CH3OH} (right panel) for the layer experiments presented in figure\,\ref{figNe: CO decroissance}. When hydrogenation of a monolayer of CO ice occurs either onto or under 1.43\,ML of Ne ice (blue and orange curves, respectively), the formation of \ce{H2CO} and \ce{CH3OH} is comparable. However, fewer products are formed compared to the hydrogenation of the monolayer of CO only. This suggests that Ne inhibits the hydrogenation of CO regardless of whether it is on top of or under the CO layer.
\begin{figure*}
    \centering
    \includegraphics[width=1\textwidth]{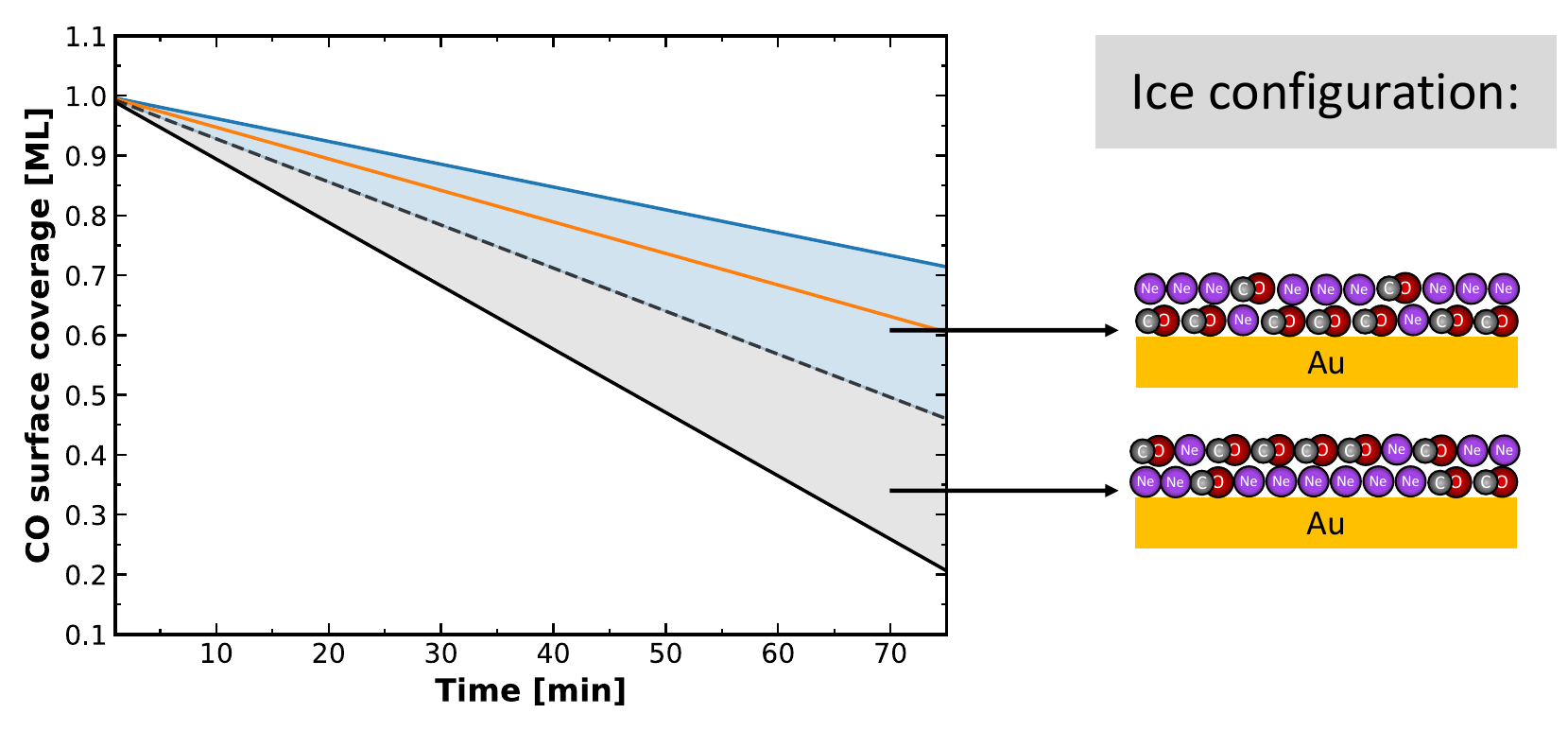}
    \caption{Simple interpretation of H bombardment on CO and Ne ice is presented in Fig.\,\ref{figNe: CO decroissance}. The blue curve shows CO ice deposited before Ne, the orange curve shows CO ice deposited after Ne, the black curve indicates the deposition of CO ice only and the dashed line represents the average slope between the blue and black curves. The blue area shows where Ne is mostly positioned above \ce{CO}, whereas the grey area indicates \ce{Ne} is mainly below \ce{CO}. The orange curve is in the blue zone, which represents the configuration where there is mostly neon on the surface.}
    \label{figC: figure_detailes_fig4_5}
\end{figure*}

Figure\,\ref{figC: figure_detailes_fig4_5} provides a simple interpretation of the layer experiments illustrated in Figure\,\ref{figNe: CO decroissance}. The blue curve corresponds to the hydrogenation of the \ce{Ne} layer deposited after the \ce{CO} layer, whereas the orange curve represents the scenario where the \ce{Ne} layer is deposited before the \ce{CO} layer. The black curve indicates the case with only a \ce{CO} layer, without \ce{Ne}. The dashed curve shows the average between the blue and black curves and delimits two distinct regions that correspond to different ice configurations. The blue area signifies a mixture of \ce{Ne} and \ce{CO}, with \ce{Ne} atoms predominantly located above the \ce{CO} layer. Conversely, the grey area shows a mixture where \ce{Ne} atoms are primarily located beneath the \ce{CO} molecules. The orange curve, which represents \ce{Ne} deposited below the \ce{CO}, falls within the blue zone, indicating that some \ce{Ne} has migrated to the top of the \ce{CO} layer and isolates some \ce{CO} from the H atoms. A comparison of Figures\,\ref{figNe: CO decroissance} and \ref{figC: figure_detailes_fig4_5} reveals that most \ce{CO} molecules are relatively inaccessible to \ce{H} atoms, even when deposited after the \ce{Ne} layer. The fact that they are not accessible to the \ce{H} atoms indicates that the \ce{CO} is below the Ne layer.

\end{appendix}

\end{document}